**Original paper**

# Ferroelectricity enhancement in ferroelectric nanotubes.

A.N. MOROZOVSKA[*], M.D. GLINCHUK and E.A. ELISEEV

Institute for Problems of Materials Science, NAS of Ukraine,

Krjijanovskogo 3, 03142 Kiev, Ukraine

Abstract

In this paper we study the size effects of the ferroelectric nanotube phase diagrams and polar properties allowing for effective surface tension and depolarization field influence. The approximate analytical expression for the paraelectric-ferroelectric transition temperature dependence on the radii of nanotube, polarization gradient coefficient, extrapolation length, surface tension and electrostriction coefficient was derived. It was shown that the transition temperature could be higher than the one of the bulk material for negative electrostriction coefficient. Therefore we predict conservation and enhancement of polarization in long ferroelectric nanotubes. Obtained results explain the observed ferroelectricity conservation and enhancement in $Pb(Zr,Ti)O_3$ and $BaTiO_3$ nanotubes.

## 1. Introduction

Ferroelectric nanoparticles of different shape are actively studied in nano-physics and nano-technology. The ferroelectric phase was studied in ferroelectric nanowires, nanotubes and nanorods [1], [2], [3], [4]. It is appeared that nanorods and nanotubes posses such polar

---

[*] Corresponding author e-mail: morozo@i.com.ua. Permanent address: V. Lashkaryov Institute of

Semiconductor Physics, NAS of Ukraine, 41, pr. Nauki, 03028 Kiev, Ukraine.

properties as remnant polarization [1] and piezoelectric hysteresis [2], [3], [4]. Moreover, co-called "confined" geometry does not destroy ferroelectric phase as predicted for spherical particles [5], [6], [7] and observed experimentally [8], but sometimes the noticeable enhancement of ferroelectric properties appears in nano-cylinders [1], [2], [3], [4].

Yadlovker and Berger [1] reported about the spontaneous polarization enhancement up to 0.25-2μC/cm$^2$ and ferroelectric phase conservation in Roshelle salt (RS) nanorods. The phenomenological description of ferroelectricity enhancement in confined nanorods has been recently proposed [9]. Morrison *et al.* [3], [4] demonstrated that PbZr$_{0.52}$Ti$_{0.48}$O$_3$ nanotubes possess perfect rectangular piezoresponse hysteresis loops.

Actually the aforementioned facts proved that the shape of nanoparticles essentially influences on the critical volume necessary for the ferroelectricity conservation possibly owing to the different depolarization field and mechanical boundary conditions [10]. In theoretical papers [5], [11] the special attention was paid to size effects, but depolarization field influence on a nanoparticle was neglected. However, depolarization field exists in the majority of confined ferroelectric systems and causes the size-induced ferroelectricity disappearance [12], [13], [14].

We suppose that a nanoparticle surface is covered with a charged layer consisted of the free carriers adsorbed in the ambient conditions [15]. The surface charges screen the surrounding medium from the nanoparticle electric field (the case of non-interacting nanoparticles assembly), but the depolarization field inside the particle is caused by inhomogeneous polarization distribution as proposed by Kretschmer and Binder [12].

For the description of nanotubes and nanowires ferroelectric properties we used the Euler-Lagrange equations, which will be solved by means of a direct variational method [13]. The approximate analytical expression for paraelectric-ferroelectric transition temperature dependence on the nanoparticle sizes, extrapolation length, effective radial stress, related to surface energy (i.e. surface tension) [16], polarization gradient and electrostriction coupling coefficients was derived.

## 2. Free energy of a nanotube

Let us consider ferroelectric cylindrical nanotube with outer radius $R_1$, inner radius $R_2$, height $h$ and polarization $P_Z(\rho, \psi, z)$ oriented along z –axes. The external electric field is $\mathbf{E} = (0, 0, E_0)$.

The bulk part of the free energy functional $\Delta G_V$ acquires the form:

$$\Delta G_V = \int\limits_{-h/2}^{h/2} dz \int\limits_{0}^{2\pi} d\psi \int\limits_{R_2}^{R_1} \rho\, d\rho \left( \frac{\alpha_R(T)}{2} P_Z^2 + \frac{\beta}{4} P_Z^4 + \frac{\gamma}{6} P_Z^6 + \frac{\delta}{2} (\nabla P_Z)^2 - P_Z \left( E_0 + \frac{E_Z^d}{2} \right) \right) \quad (1)$$

Material coefficients $\delta > 0$ and $\gamma > 0$, while $\beta < 0$ for the first order phase transitions or $\beta > 0$ for the second order ones. The coefficient $\alpha_R(T)$ in Eq.(1) should be renormalized by the external stress (see e.g. Ref. [14], [17], [18])

$$\alpha_R(T, R_1, R_2) = \alpha_T(T - T_C) - 2Q_{12}\sigma(R_1, R_2) \quad . \quad (2a)$$

Here parameters $T_C$ and $Q_{12}$ are respectively Curie temperature and electrostriction coefficient of the bulk material, $\alpha_T$ is proportional to the inverse Curie constant. The stress $\sigma(R_1, R_2)$ is caused by the radial pressures $p_1$ and $p_2$ (see Ref. [18]):

$$\sigma(R_1, R_2) \approx \frac{1}{1 - (R_2/R_1)^2} \left( p_1 + \left( \frac{R_2^2}{R_1^2} \right) p_2 \right) \exp\left( -\frac{\Delta R}{R_1 - R_2} \right) \quad (2b)$$

Parameter $\Delta R$ is the characteristic thickness of a nanotube, below which the factor $\frac{1}{1 - (R_2/R_1)^2}$ becomes too high; thus it characterizes the stress relaxation via dislocation appearance [19]. Hereinafter we assume that radial pressures $p_1 = -\mu_1/R_1$ and $p_2 = -\mu_2/R_2$ is caused by the surface energy excess and put surface tension coefficients [8], [20] equal $\mu_{1,2} = \mu > 0$. In the case $\sigma(R_1, R_2) = -\dfrac{\mu}{R_1 - R_2}\exp\left(-\dfrac{\Delta R}{R_1 - R_2}\right)$ and so it reaches maximum at

$R_1 - R_2 = \Delta R$, where $\Delta R = \mu/\sigma_{\max}$.

The exact expression for depolarization field $\mathbf{E}^d(\rho, \psi, z)$ inside the cylindrical

nanoparticle covered with screening charges is derived in Ref. [18]. Its estimation for a thick

tube ($R_1 - R_2 >> \sqrt{\delta}$) has the form:

$$E_Z^d(\rho, \psi, z) \approx -\frac{4\pi}{1 + (k_{01}h/2\pi R_1)^2}\left(P_Z(\rho, \psi, z) - \frac{2}{h}\int_{-h/2}^{h/2}dz P_Z(\rho, \psi, z)\right) \qquad (3)$$

Hereinafter $k_{01}(R_1, R_2)$ is the lowest root of the equation

$J_0(k_{01}R_2/R_1)N_0(k_{01}) - J_0(k_{01})N_0(k_{01}R_2/R_1) = 0$ ($J_0(x)$ and $N_0(x)$ are Bessel and Neiman

functions of zero order respectively).

The surface part of the polarization-dependent free energy $\Delta G_S$ acquires the form:

$$\Delta G_S = \delta \int_0^{2\pi} d\psi \left(\int_{R_2}^{R_1}\frac{\rho}{\lambda_b}d\rho\left(P_Z^2\left(z = \frac{h}{2}\right) + P_Z^2\left(z = -\frac{h}{2}\right)\right) + \int_{-h/2}^{h/2}dz\left(\frac{R_1}{\lambda_S}P_Z^2(\rho = R_1) + \frac{R_2}{\lambda_S}P_Z^2(\rho = R_2)\right)\right)$$
$$(4)$$

We introduced longitudinal and lateral extrapolation lengths $\lambda_b \neq \lambda_S$ in Eq. (4). Hereinafter we regard these extrapolation lengths positive.

Variation of the free energy expression $\Delta G = \Delta G_V + \Delta G_S$ yields the following Euler-

Lagrange equations:

$$\begin{cases} \alpha_R P_Z + \beta P_Z^3 + \gamma P_Z^5 - \delta\left(\dfrac{\partial^2}{\partial z^2} + \dfrac{1}{\rho}\dfrac{\partial}{\partial\rho}\rho\dfrac{\partial}{\partial\rho} + \dfrac{1}{\rho^2}\dfrac{\partial^2}{\partial\psi^2}\right)P_Z = E_0 + E_Z^d(\rho, \psi, z), \\ \left(P_Z + \lambda_b\dfrac{dP_Z}{dz}\right)\Bigg|_{z=h/2} = 0, \qquad \left(P_Z - \lambda_b\dfrac{dP_Z}{dz}\right)\Bigg|_{z=-h/2} = 0, \qquad (5) \\ \left(P_Z + \lambda_S\dfrac{dP_Z}{d\rho}\right)\Bigg|_{\rho=R_1} = 0, \qquad \left(P_Z - \lambda_S\dfrac{dP_Z}{d\rho}\right)\Bigg|_{\rho=R_2} = 0, \end{cases}$$

The polarization distribution in the ferroelectric phase should be found by direct variational method [13] allowing for possible polydomain states appearance in confined particles. Briefly, the domain wall energy is represented by the correlation term $\sim \delta(\nabla P_Z(\rho, z))^2$ in Eq.(1), so polydomain states could be studied with the help of the free energy (1)-(4). It is appeared that single-domain state is energetically preferable for infinite tubes and wires, since depolarization field is absent and correlation energy is minimal for single-domain case.

## 3. Phase diagram of the long nanotubes

We derived [18] the interpolation for the paraelectric-ferroelectric transition temperature $T_{CR}(R_1, R_2)$ of the long nanotubes ($h >> R_1$ so $E_d \to 0$):

$$T_{CR}(R_1, R_2) = T_C - \frac{2Q_{12}\mu}{\alpha_T(R_1 - R_2)}\exp\left(-\frac{\Delta R}{R_1 - R_2}\right) - \delta\frac{k_{01}^2(R_1, R_2)}{\alpha_T R_1^2}, \qquad (6)$$

The inequality $(\lambda_S/R_1) << 1$ used in Eq.(6) is valid for typical extrapolation lengths $\lambda_S = 0.3...5\,\text{nm}$ and radiuses $R_1 = 30...500\,\text{nm}$.

The first term in Eq. (6) is the bulk transition temperature, the second term is related to the coupling of radial stress with polarization via electrostriction effect, the third term is caused by correlation effects. The correlation term is always negative and thus only decreases the transition temperature, whereas the electrostriction term depends on the $Q_{12}$ sign, however $Q_{12} < 0$ for most of the perovskite ferroelectrics.

Let us make some estimations of the second and third terms in Eq.(6) for perovskites BaTiO$_3$ and PbTiO$_3$. Using parameters $Q_{12} = -0.043\,\text{m}^4/\text{C}^2$, $T_C = 400\,\text{K}$ (BaTiO$_3$) and $Q_{12} = -0.046\,\text{m}^4/\text{C}^2$, $T_C = 666\,\text{K}$ (PbZr$_{0.5}$Ti$_{0.5}$O$_3$); $\mu_{1,2} = 5 - 50\,\text{N/m}$ (see e.g. Ref. [8]) and $\delta = 10^{-19} - 10^{-17}\,\text{m}^2$, we obtained that $\left|\dfrac{2\mu Q_{12}}{\alpha_T T_C}\right| \approx 2 - 17\,\text{nm}$, $\sqrt{\dfrac{\delta}{\alpha_T T_C}} \approx 2.3 - 23\,\text{nm}$ for BaTiO$_3$

and $\left|\dfrac{2\mu Q_{12}}{\alpha_T T_C}\right| \approx 3 - 26\ \text{nm}$, $\sqrt{\dfrac{\delta}{\alpha_T T_C}} \approx 1.9 - 19\ \text{nm}$ for PbZr$_{0.5}$Ti$_{0.5}$O$_3$ respectively. So both terms are comparable with unity at nanoparticle radius $\sim$ 2-25 nm.

Taking into account that the gradient coefficient $\sqrt{\delta} \sim 0.3...3\ \text{nm}$, i.e. it is of several lattice constants, we introduced the parameters and dimensionless variables that correspond to the lattice constants units $R_\mu = \dfrac{2\mu Q_{12}}{\alpha_T T_C \sqrt{\delta}}$, $R_S = \sqrt{\dfrac{1}{\alpha_T T_C}}$, $r_{1,2} = \dfrac{R_{1,2}}{\sqrt{\delta}}$, $\Delta r = \dfrac{\Delta R}{\sqrt{\delta}}$. So, the parameter $R_\mu$ is negative for most of perovskites with $Q_{12} < 0$. In accordance with our estimations we obtained that $R_S \sim 5...10$ and $|R_\mu| \sim 8...80$.

In Fig.1 we present phase diagrams calculations based on the Eqs.(6). It is clear from

Figs.1 that the critical radius significantly depend on the nanotube thickness, namely the

critical radius is smallest for nanowires ($r_2 = 0$), slightly bigger for "thick" nanotubes

($r_2/r_1 < 0.1$) and biggest the "thin" ones ($r_2/r_1 \approx 1$). The transition temperature $T_{CR}(r_1, r_2)$

tends to the bulk value $T_C$ at $r_1 \to \infty$ for any shape, as it should be expected for the bulk ferroelectric material.

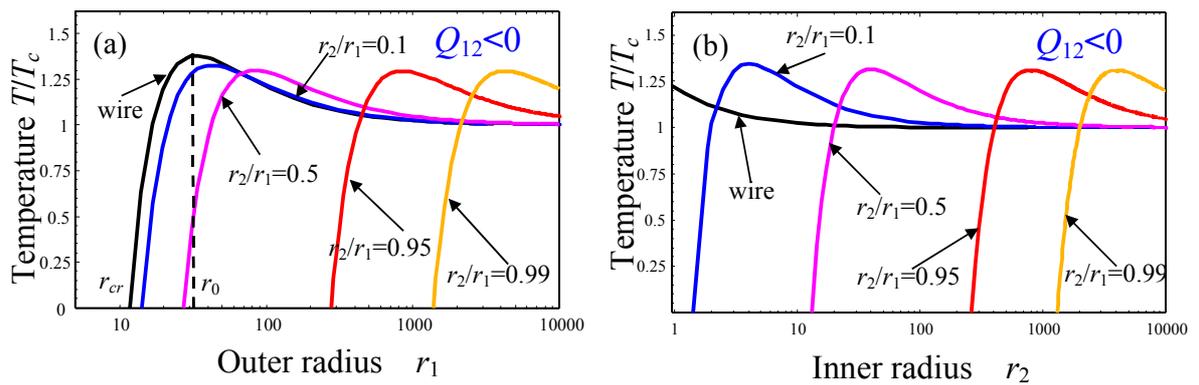

FIG.1 (Color online) Transition temperature $T_{CR}(r_1, r_2)$ vs. outer radius $r_1$ (a) and inner radius $r_2$ (b) for different ratios $r_2/r_1$ (figures near the curves). Other parameters: $\alpha_T = 2.95 \cdot 10^{-5}$, $T_C = 666\ \text{K}$, $R_S \approx 7$,

$\Delta r = 5$ and $R_\mu = \pm 25$ correspond to PbZr$_{0.5}$Ti$_{0.5}$O$_3$.

At $R_\mu < 0$ nanowires and nanotubes reveal noticeable increase of transition temperature ($T_{CR}/T_C > 1$) in the vicinity of the optimal radius $r_0$ (competition between the radial stress and the correlation effect). Note, that the polarization enhancement in thin nanotubes could be explained by peculiarities of the stress size dependence given by Eq.(2b).

## 4. Polarization enhancement in the long nanotubes

For long enough nanotubes we derived approximate analytical expressions for the free energy with renormalized coefficients, namely:

$$\Delta G \approx \frac{\alpha_T}{2}(T - T_{CR}(R_1, R_2))P_n^2 + \frac{\beta}{4}P_n^4 + \frac{\gamma}{6}P_n^6 - P_n E_0. \qquad (7)$$

Here $T_{CR}(R_1, R_2)$ is given by Eq.(6). The free energy (7) has conventional form of power series on the averaged polarization. Thus, one immediately obtains the average values after solving algebraic equations. Namely, for the ferroelectrics with the second order phase transition: spontaneous polarization $P_{nS} = \sqrt{-\alpha_T(T - T_{CR}(R_1, R_2))/\beta}$ and thermodynamic coercive field $E_C^n = \frac{2P_{nS}}{3\sqrt{3}}\alpha_T(T - T_{CR}(R_1, R_2))$. The dependences of spontaneous polarization $P_{nS}$ and thermodynamic coercive field $E_C^n$ on the nanotube outer radius are depicted in Fig.2.

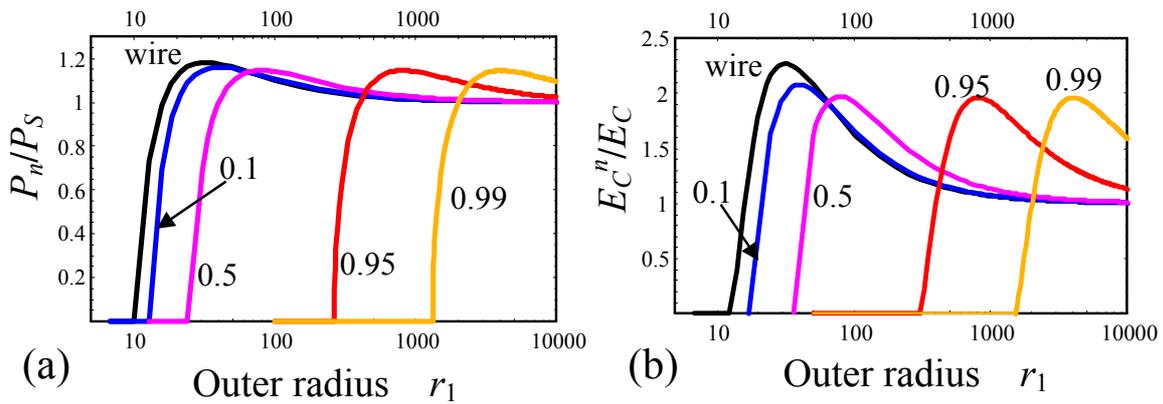

FIG.2 (Color online) Spontaneous polarization $P_{nS}(T)/P_S(T)$ (a) and thermodynamic coercive field $E_C^n(T)/E_C(T)$ (b) vs. outer radius $r_1$ for different ratios $r_2/r_1$ (figures near the curves). Other parameters:

$\alpha_T = 2.95 \cdot 10^{-5}$, $T = 300$ K, $T_C = 666$ K, $R_S \approx 7$, $\Delta r = 5$ and $R_\mu = -25$ correspond to $PbZr_{0.5}Ti_{0.5}O_3$.

It is clear from the figure that the regions with spontaneous polarization $P_{nS}$ higher than its bulk value $P_S(T)$ always exist. In the same region or radiuses the coercive field $E_C^n$ is higher than the bulk value $E_C(T)$. The coercive field firstly increases with the tube outer radius increase, quickly reaches the maximum and then decreases tending to the bulk value with the tube outer radius increase.

## 5. Comparison with experiment

Polar properties enhancement in confined RS nanorods was reported by Yadlovker and Berger [1] and partly explained earlier [9].

Recently Morrison et al. [3], [4] demonstrated that long $Pb(Zr,Ti)O_3$ and $BaTiO_3$ nanotubes posses perfect piezoelectric properties. Appeared that measured effective piezoelectric response value $d_{33}^{eff} = u_3/U$ ($u_3$ is a surface displacement, $U$ is the voltage applied to the AFM probe) is close or higher than the bulk ones. The piezoresponse of the uniformly polarized cylindrical domains tube is considered in Ref. [21]:

$$d_{33}^{eff} = t_{13}(R_1, R_2, \gamma)d_{31} + t_{51}(R_1, R_2, \gamma)d_{15} + t_{33}(R_1, R_2, \gamma)d_{33} \ . \qquad (8)$$

Where rather cumbersome functions $t_{13}(R_1, R_2, \gamma)$ depend only on tube radiuses, dielectric anisotropy coefficient $\gamma = \sqrt{\varepsilon_{33}/\varepsilon_{11}}$ and probe electric field distribution, $d_{33} = 2Q_{11}\chi_{33}P_3$, $d_{31} = 2Q_{12}\chi_{33}P_3$ and $d_{15} = 2Q_{44}\chi_{11}P_3$ in the case for a rigid model for polarization $P_Z \equiv P_S$. However, the relation $d_{ij}^{eff} \sim \chi_{kj}P_Z(\rho, \psi)$ is not rigorous for the definite distributions of polarization $P_Z(\rho, \psi)$ and susceptibility $\chi_{33}(\rho, \psi)$. However, it is obvious, that $d_{33}^{eff}(U) \sim P_n(U)(\chi_{33}(U) + \vartheta)$, here $\vartheta \sim d_{51}\chi_{11}/d_{33}$ is a fitting parameter. We compare the piezoresponse loop shape obtained for $PbZr_{52}Ti_{48}O_3$ nanotube [4] with our modelling in Fig.3.

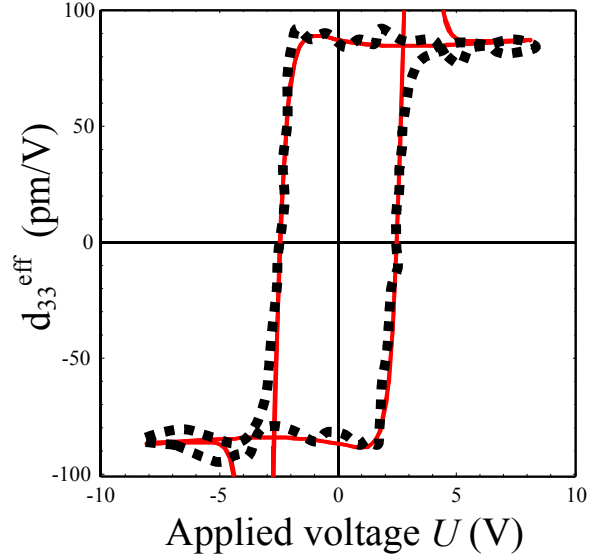

FIG.3 (Color online) Effective piezoresponse $d_{33}^{eff}$ of $PbZr_{52}Ti_{48}O_3$ nanotube (outer diameter 700nm, wall thickness 90nm, length about 30μm) vs. applied voltage $U$; the loop was centered. Squares are experimental data of Morrison et al. [4] solid curve is our fitting for $R_1 \approx 700$, $R_2 \approx 610$ (i.e. $\sqrt{\delta} = 1\,\mathrm{nm}$), $\alpha_T = 2.95 \cdot 10^{-5}$, $T_C = 666\,\mathrm{K}$, $T = 300\,\mathrm{K}$, $R_S \approx 7$, $\Delta r = 5$, $R_\mu = -5$, $\vartheta = 0.25$.

Despite aforementioned remarks our fitting is in a surprisingly good agreement with observed local piezoresponse hysteresis loops.

We obtained, that the possible reason of the polar properties enhancement in confined ferroelectric nanotubes and nanowires is the radial stress coupled with polarization via electrostriction effect under the decrease of depolarization field for long cylindrical nanoparticles.